\documentclass[pra,aps,showpacs,groupedaddress,superscriptaddress,twocolumn, numerical]{revtex4-1}

\usepackage[utf8]{inputenc}
\usepackage{color}
\usepackage{bbm} 
\usepackage{nicefrac}
\usepackage{amsfonts,amsmath,amssymb,stmaryrd}
\usepackage{braket}

\usepackage[autostyle=true]{csquotes}

\usepackage{tabularx}
\usepackage{multirow} 
\usepackage{hhline}
\usepackage{graphicx}
\usepackage{subfigure}  % use for side-by-side figures
\usepackage{bbm} 
\usepackage{hyperref}
\usepackage[pdftex]{epsfig}
\usepackage{mathrsfs}
\usepackage{verbatim}
\usepackage{centernot}
\usepackage{array}
\usepackage{cancel}
\usepackage{ifthen}
\usepackage{bm}	% bold in math mode
\usepackage{todonotes}
\presetkeys{todonotes}{size=\tiny}{}

\InputIfFileExists{gitinfo-latex.inc}{}{}

\usepackage[acronym,nomain]{glossaries}

\usepackage{units}
\usepackage{upgreek}

\begin{document}
	\title{A Single Atom Thermometer for Ultracold Gases}
	
	\author{Michael Hohmann}
	\affiliation{Department of Physics and Research Center OPTIMAS, University of Kaiserslautern, Germany}
	
	\author{Farina Kindermann}
	\affiliation{Department of Physics and Research Center OPTIMAS, University of Kaiserslautern, Germany}
	
	\author{Tobias Lausch}
	\affiliation{Department of Physics and Research Center OPTIMAS, University of Kaiserslautern, Germany}
	
	\author{Daniel Mayer}
	\affiliation{Department of Physics and Research Center OPTIMAS, University of Kaiserslautern, Germany}
	\affiliation{Graduate School Materials Science in Mainz, Gottlieb-Daimler-Strasse 47, 67663 Kaiserslautern, Germany}
	
	\author{Felix Schmidt}
	\affiliation{Department of Physics and Research Center OPTIMAS, University of Kaiserslautern, Germany}
	\affiliation{Graduate School Materials Science in Mainz, Gottlieb-Daimler-Strasse 47, 67663 Kaiserslautern, Germany}
	
	\author{Artur Widera}
	\affiliation{Department of Physics and Research Center OPTIMAS, University of Kaiserslautern, Germany}
	\affiliation{Graduate School Materials Science in Mainz, Gottlieb-Daimler-Strasse 47, 67663 Kaiserslautern, Germany}

	
	\date{\today}
	
	\begin{abstract}
		We use single or few Cs atoms as thermometer for an ultracold, thermal Rb cloud. 
		Observing the thermometer atoms' thermalization with the cold gas using spatially resolved fluorescence detection, we find an interesting situation, where a fraction of thermometer atoms thermalizes with the cloud while the other fraction remains unaffected.
		We compare release-recapture measurements of the thermometer atoms to Monte-Carlo simulations while correcting for the non-thermalized fraction, and recover the cold cloud's temperature.
		The temperatures obtained are verified by independent time-of-flight measurements of the cold cloud's temperature.
		We also check the reliability of our simulations by first numerically modelling the unperturbed in-trap motion of single atoms in absence of the cold cloud, and second by performing release-recapture thermometry on the cold cloud itself.
		Our findings pave the way for local temperature probing of quantum systems in non-equilibrium situations.
	\end{abstract}	
	
	\maketitle
	
\section{Introduction}
	Thermometry of quantum gases has recently attracted much interest especially for non-equilibrium situations~\cite{Gring2012,Hofferberth2007,Kinoshita2006}.
	An interesting question regards the local state of a system. An appealing option to experimentally address this question is the use of single atoms as thermometers immersed in the ultracold gas, which can be understood as the most extreme case of minority species thermometry \cite{Olf2015, McKay2010, Regal2007, Nascimbene2010, Spiegelhalder2009}.
	In order to serve as a local temperature probe, the particle should, first, thermalize much faster with its local environment than the many-body system does globally.
	Second, detection of the probe atoms should feature high spatial resolution.
	And third, the probe should not affect the temperature of the many-body system.
	This suggests using few or even single particles having a much larger inter-species interaction, hence thermalization rate with atoms of the gas, than the intra-species interaction of the gas atoms.
	We realize such a system by using single $^{133}$Cs atoms as thermometers for a cold, thermal $^{87}$Rb cloud, where the inter-species scattering length is $\approx6.3$ times larger than the Rb-Rb intra-species scattering length \cite{Lercher2011,vanKempen2002} and may also be tuned by employing a Feshbach resonance \cite{Takekoshi2012}.
	The determination of a single thermometer atom's temperature relies on measuring the energy distribution of an average over many identical realizations.
	For Maxwell-Boltzmann distributed energies, a temperature can be assigned.
	While the energy distribution of tightly confined particles may be measured via vibrational spectroscopy \cite{Foerster2009,Diedrich1989}, we focus on a complementary method which can be applied to weakly confined particles, where the vibrational states within the external potential are too small to be resolved.
	This so-called release-recapture method detects the remaining fraction of trapped atoms after a short release from the trap for a variable period of time \cite{Reymond2003,Mudrich2002}.
	The data is compared to Monte-Carlo simulations assuming different initial temperatures to identify the simulation temperature reproducing the experimental data.
	
	Experimentally, the cold Rb cloud is prepared in a dipole trap, the single Cs atoms are transferred to the same potential and after an evolution time, a release-recapture experiment is performed, similar to our previous work \cite{Spethmann2012}.
	For various temperatures between 1 and $\unit[5]{\upmu K}$, we measure the thermometer atoms' spatial distribution and observe the atoms separating into two distinct fractions.
	The first fraction thermalizes with the cold Rb cloud, whereas the second fraction does not and oscillates freely in the trap.
	Hence, information about the cloud temperature is only encoded in the first, thermalized fraction.
	Performing release-recapture experiments on the thermometer atoms with and without a cold cloud enables us to account for the non-thermalized fraction's influence. Consequently, we retrieve the cloud's temperature, in quantitative agreement to independent time-of-flight measurements of the cloud.
	We further show the reliability of our method and its applicability to different scenarios, by performing release-recapture thermometry directly with the cloud, again in quantitative agreement to time-of-flight thermometry.
	
\section{Experimental System}	
\begin{figure}[tb]
	\begin{center}
		\includegraphics{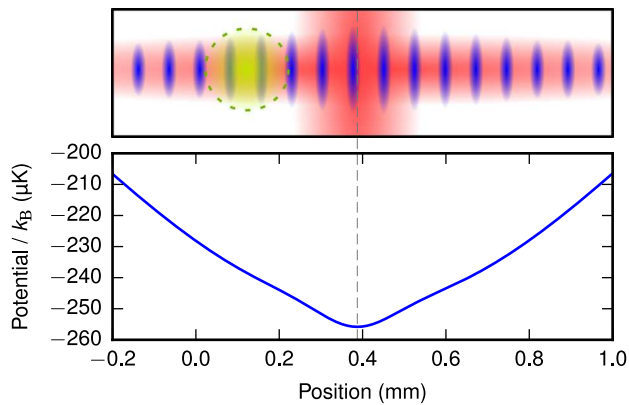}
	\end{center}
	\caption{Schematic drawing of the involved traps and the calculated dipole potential for Cs used for the measurement discussed in section~\ref{sec:movie}. The single atom MOT (yellow, dashed) is located left of the dipole trap crossing region (red). A conveyor-belt lattice (blue) is used to freeze the axial motion of the single atoms during fluorescence imaging. The vertical dashed line indicates the potential minimum.}
	\label{fig:dipole_potential_movie_drawing}
\end{figure}		
We prepare cold, thermal clouds with typically $1 \times 10^4 \dots 5 \times 10^4$ Rb atoms in the $\ket{F=1, m_F=0}$ state and temperatures between 0.1 and $\unit[5]{\upmu K}$ in a crossed dipole trap at \unit[1064]{nm}.
The traps involved in the preparation of both species are schematically shown in Fig.~\ref{fig:dipole_potential_movie_drawing}.
For experiments with Cs, the trap frequencies are typically $\omega_\textrm{r} = 2 \pi \times \unit[1.8]{kHz}$ radial to the horizontal beam and $\omega_\textrm{ax} = 2 \pi \times \unit[60]{Hz}$ in axial direction, nearly equal for both species.
The trap depth is $\unit[260]{\upmu K} \times k_\textrm{B}$ ($\unit[150]{\upmu K} \times k_\textrm{B}$) for Cs (Rb), leading to Rb cloud densities on the order of $10^{12}$~-~$\unit[10^{13}]{cm^{-3}}$.

For the thermometer atoms, up to $\approx6$ Cs atoms are trapped from the background gas in a high gradient magneto-optical trap (MOT), located approximately $ \unit[300]{\upmu m}$ away from the potential minimum of the dipole trap.
Released from the MOT, the Cs atoms move toward the Rb cloud at the trap center, where they serve as thermometers.
To measure the Cs atoms' spatial distribution, we ramp up a one-dimensional optical conveyor-belt lattice along the horizontal dipole trap beam to freeze the atomic position in one dimension.
The lattice laser has a wavelength of $\lambda_\textrm{lat} = \unit[790]{nm}$, blue-detuned for Cs with a maximum potential of $\unit[830]{\upmu K} \times k_\textrm{B}$.
To avoid light-assisted Cs loss \cite{ThesisNicolasSpethmann, John2010, Weber2010, Weiner1999} during final imaging, the Rb cloud is pushed out of the trap with near-resonant light.
Subsequently, the horizontal dipole trap depth for Cs is increased to $\unit[1480]{\upmu K} \times k_\textrm{B}$ and an optical molasses is switched on for Cs. 
Fluorescence detection allows to image the Cs atoms with a resolution of $\unit[1.6]{\upmu m}$, revealing their spatial distribution. This high resolution imaging is restricted to a field of view of $\unit[246]{\upmu m}$, much smaller than the length scale at which relevant dynamics take place.
We therefore extend our effective field of view by shifting distant parts of the lattice into the imaging region and stitch the individual partial images.
If we are only interested in the number of Cs atoms, a different, faster scheme is employed:
Without using the optical lattice, Rb is pushed out of the dipole trap, the Cs atoms are transferred back to the MOT and their fluorescence level is used to determine the atom number \cite{Haubrich1996,Ruschewitz1996}.
A more detailed description of our experimental setup is given in \cite{Hohmann2015}.

\section{Release-Recapture Procedure and Simulations}
We measure the atomic temperature with a release-recapture procedure, that is both applicable to the cold Rb cloud and single Cs atoms, but the experimental details vary.
In this section, we focus on the release-recapture procedure for the Cs thermometer atoms.
For a release-recapture measurement, the dipole trap is switched off for a short, variable release time.
During the release, the spatial distribution of the atoms expands due to their thermal energy and the atoms fall freely, due to gravity.
A short push-out pulse for Rb consisting of resonant cooler and repumper light is applied \unit[200]{$\upmu$s} after the release, which is advantageous for two reasons:
First, the chance of collisions between Cs and already accelerated Rb atoms is minimized, since the Rb density continually decreases during release.
Second, the probability of light-assisted Cs loss during the pulse is minimized because even low momentum transfer can separate the two species rapidly, as there is no trapping force.
After recapture, a release-time dependent fraction of atoms have enough energy to escape the trap.
The probability for an atom escaping the trap, however, does not only depend on its total energy but on the distribution of that energy to the atom's degrees of freedom.
After a  typical waiting time of \unit[50]{ms} for losses to occur, the recaptured atoms are transferred to the MOT, where their number is detected.
From the fraction of remaining atoms as a function of release duration ('release-recapture curve'), the atoms' initial temperature can be deduced.
Although there is an approximative formula \cite{Mudrich2002} to predict the release-recapture curve for a given temperature, the approximations, e.g.~neglecting gravity, do not apply to our case.
Therefore, we employ a classical Monte-Carlo method to simulate the atoms' dynamics from their release to the detection:
The atoms are modeled as point masses that do not interact with each other and are exposed to gravity and dipole trap forces.
A 3rd order Runge-Kutta method is used to solve the equations of motion.
Representative initial phase space coordinates for the thermometer atoms are obtained numerically by an iterative process:
We start with $1000$~atoms located at the trapping potential minimum.
In each iteration step, new, Maxwell-Boltzmann distributed velocities for the desired temperature are assigned to the atoms and their phase space coordinates are evolved for a time, short compared to the trap oscillation period.
The assignment of new velocities in each step effectively models a contact to a thermal bath, compensating the transition of thermal to potential energy as the position distribution expands to its equilibrium size during the time evolution.
The process is repeated until the position distribution has converged. The final sample of trapped thermal atoms represents the statistical velocity and position distributions.
This sample's phase-space coordinates are time-evolved to the end of the waiting time for each release time. Subsequently, the atoms remaining in the trapping region are counted, yielding a prediction for a release-recapture curve for the given temperature.
The simulation relies on precise knowledge of the trap geometry and its time dependence.
Therefore, the trap geometry is determined with the help of Rb trap frequency measurements. Furthermore, absorption pictures of the Rb cloud are taken to determine where the two dipole trap beams intersect with respect to the horizontal beam's focus position.
Also, the time-dependence of the beam powers are recorded for each release-time, such that delays and ramp-response functions are included in the simulation.

\section{Single Atom Dynamics and Position-resolved Thermalization Dynamics}
\label{sec:movie}
\begin{figure}[tb]
	\begin{center}
		\includegraphics{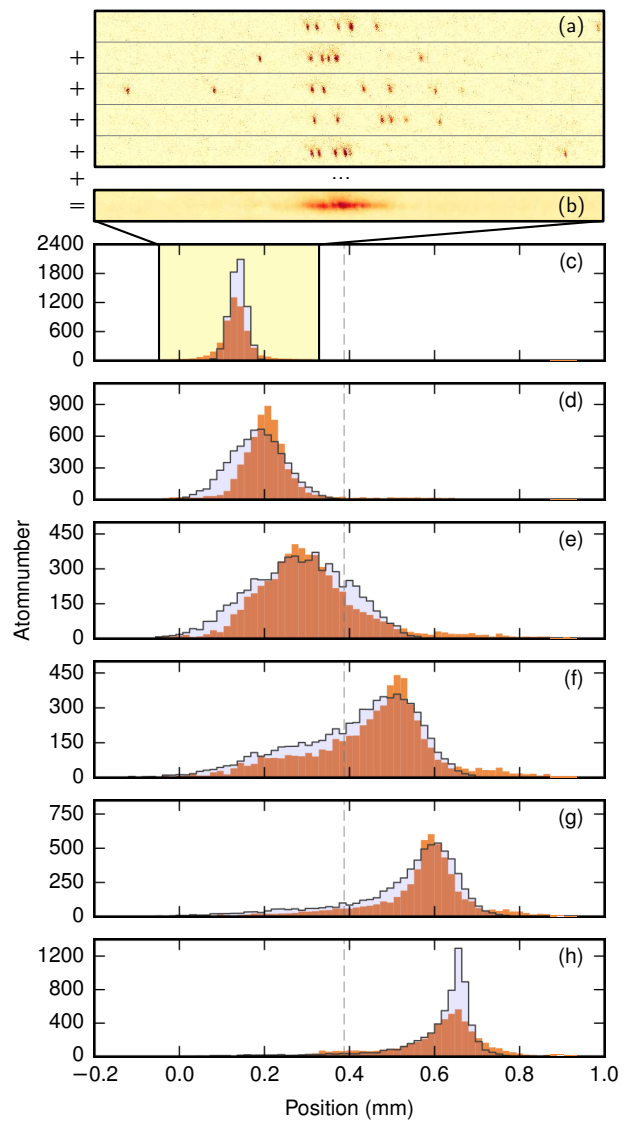}
	\end{center}
	\caption{Dynamics of single Cs atoms in the crossed dipole trap. Many fluorescence images of single or few Cs atoms (a) are taken at a time $t$ after extinguishing the MOT. The accumulated signal for $t=\unit[5]{ms}$ is shown in (b). A quantitative analysis is given in the histograms (c)-(h), where the experimentally obtained atom positions (orange, borderless) are shown together with Monte-Carlo simulation (light-blue, solid border) for $t=\unit[5,9,\dots,25]{ms}$. The simulation time starts with a delay of \unit[3.3]{ms}, taking account for the slowly decaying magnetic field of the MOT. Maximum overlap between experiment and simulated data is obtained for a MOT located at $\unit[0.133]{mm}$ at a temperature of \unit[1.9]{$\upmu$K}. The vertical dashed line indicates the position of the potential minimum.}
	\label{fig:movie}
\end{figure}

Since we are interested in the temperature of individual Cs thermometer atoms thermalized with the Rb gas, it is important to precisely know the time dependent spatial distribution of thermometer atoms in the trap.
This allows identifying, for example, the onset of thermalization when the atoms find first contact with the cloud after extinguishing the MOT.
More importantly, it allows determining the time of maximum separation between the thermalized and the non-thermalized fractions of Cs atoms.
First, we record the Cs dynamics without Rb by extinguishing the MOT and switching on the lattice potential after a variable evolution time.
The dynamics in Fig. \ref{fig:movie} show quasi-harmonic center-of-mass oscillations, while the dispersion of the position distribution originates from the atoms' thermal energy and the anharmonicity of the trap.
We fit the data with a classical Monte-Carlo simulation and find good agreement for a Cs temperature of \unit[1.9]{$\upmu$K}, consistent with \cite{Salomon1990} for our MOT parameters.
Thus, the Cs atoms occupy a non-thermal state in a $\Delta E_\mathrm{kin}=\unit[1.9]{\upmu K} \times k_\textrm{B}$ wide energy band of axial vibration states at the potential energy of their initial position $E_\mathrm{pot}=\unit[16]{\upmu K} \times k_\textrm{B}$.
In this simulation, the Cs atoms initially have Maxwell-Boltzmann distributed velocities and are normally distributed around $z=\unit[0.133]{mm}$ with standard deviations of $\sigma_r = \unit[7.6]{\upmu m}$ in radial direction and $\sigma_z = \nicefrac{\sigma_r}{2}$ in the observed, axial direction.

\begin{figure}[tb]
	\begin{center}
		\includegraphics{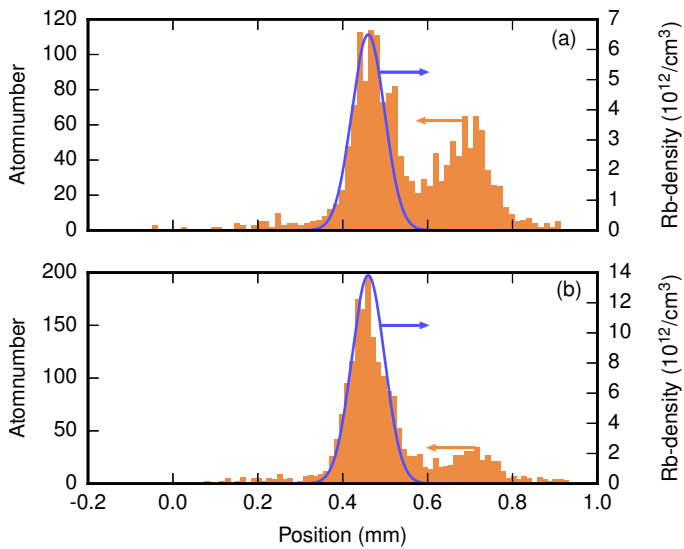}
	\end{center}
	\caption{Spatial density distribution of Cs (orange bars) with Rb at $\unit[2.2]{\upmu K}$ (blue line) with $6.6 \times 10^{3}$ (a) and $14 \times 10^{3}$ atoms (b) being present in the crossing region of the trap at the time when the non-thermalized fraction of Cs reaches the right turning point in the trap. The fraction of non-thermalized Cs is  decreased in (b) due to a higher density of Rb, thus higher effective cross-section for collisions.}
	\label{fig:density_variation}
\end{figure}		

The dynamics strongly change when a Rb cloud is present in the crossing region of the dipole trap, as shown in Fig.~\ref{fig:density_variation}:
A fraction of Cs atoms localizes at the position of the cloud, as soon as they get in contact with Rb.
The transition of these atoms to the thermalized state appears quasi-instantaneously, suggesting a fast thermalization.
This is consistent with high collision-rates of 3.1~-~\unit[5.4]{kHz} at the examined peak Rb densities of $5.5 \times 10^{12} \ldots \unit[1.2 \times 10^{13}]{cm^{-3}}$, caused by a relatively large inter-species scattering length of $a_\textrm{RbCs} \approx 630 a_\textrm{B}$ \cite{Lercher2011} with $a_\textrm{B}$ the Bohr radius.
While these atoms may serve as thermometers for the cloud, the complementing fraction of Cs atoms oscillates unperturbed to the opposite side of the crossing region.
This non-thermalized fraction depends on the Rb cloud's size and density:
Fig.~\ref{fig:density_variation} exemplarily shows spatial Cs distribution measurements for Rb clouds at $\unit[2.2]{\upmu K}$ and atom numbers of $6.6\times10^3$ and $14\times10^3$, where $45\%$ and $25\%$ of the Cs atoms, respectively, continue oscillating.

\section{Single Atom Thermometry}

\begin{figure}[tb]
	\begin{center}
		\includegraphics{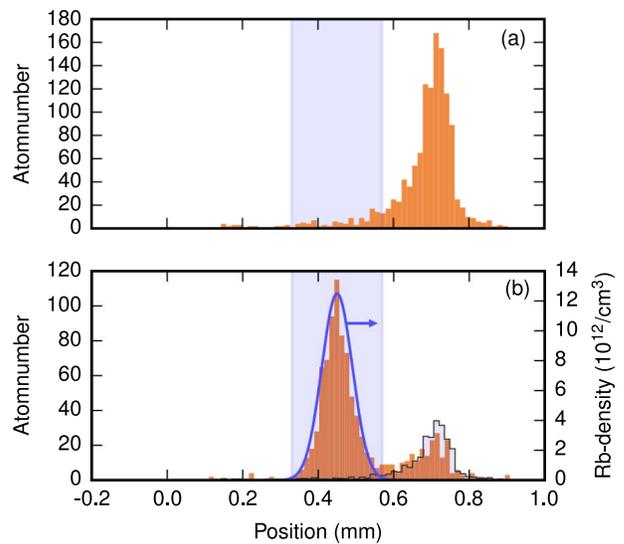}
	\end{center}
	\caption{Dynamics of single Cs atoms (orange, borderless) without Rb (a) and with Rb (b) at an evolution time of \unit[27]{ms}, when unperturbed Cs is at a turning point of its axial oscillation. The Rb density profile is shown as a solid blue line and matches the spatial distribution of the thermalized fractions, since the trap frequencies are almost equal for the two species. From the fraction of atoms outside of the blue shaded area in (b), we compute the total fraction of non-thermalized Cs atoms (light-blue, solid border) with the help of (a). In this measurement, \unit[27]{\%} of the Cs atoms are not thermalized with Rb and exhibit an unperturbed distribution.}
	\label{fig:movie_fraction_analysis}
\end{figure}		
\begin{figure}[tb]
	\begin{center}
		\includegraphics{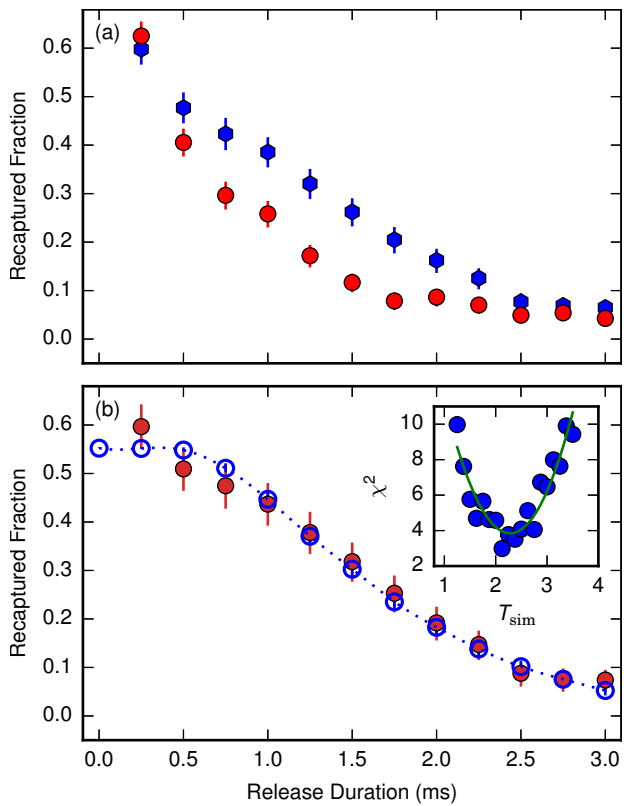}
	\end{center}
	\caption{Release-recapture thermometry on the thermalized fraction of single Cs thermometer atoms. Release-recapture curves for Cs with Rb being present in the trap (a, blue octagons) and the correction curve for Rb being absent (a, red circles) at an evolution time of \unit[27]{ms} are shown. An effective release-recapture curve of the thermalized fraction ((b), filled red circles) is computed as the difference between the two curves weighted with the thermalized and non-thermalized fraction's size. From the Monte-Carlo simulations we determine a release-recapture temperature of $T_{\textrm{rr}}=\unit[2.3 \pm 0.5]{\upmu K}$ (empty blue circles), in good agreement with the Rb time-of-flight temperature $T_{\textrm{tof}}=\unit[2.15 \pm 0.15]{\upmu K}$. Inset of (b): The $\chi^2$ values for simulated release-recapture curves at initial temperatures $T_\textrm{sim}$ follow a parabolic shape, while the parabola's slope yields the uncertainty of $T_{\textrm{rr}}$ as described in the text.}
	\label{fig:Cs_RR}
\end{figure}		
\begin{figure}[tb]
	\begin{center}
		\includegraphics{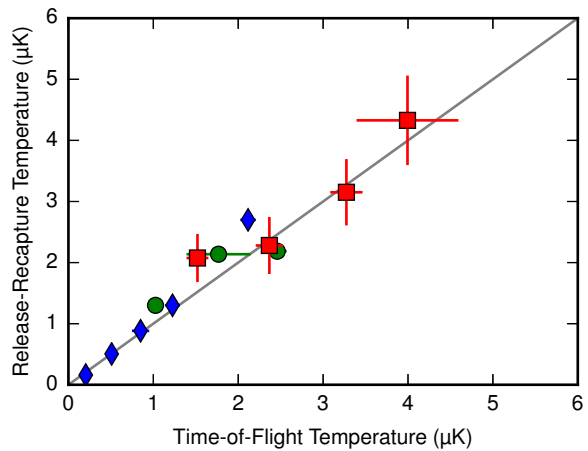}
	\end{center}
	\caption{Comparison of release-recapture temperatures $T_\textrm{rr}$ to independently measured time-of-flight temperatures $T_\textrm{tof}$, with 1$\sigma$ confidence intervals. 
		Single-atom thermometry (red squares): The temperatures $T_\textrm{rr}$ of the thermalized fraction of thermometer atoms agree well with the time-of-flight temperatures $T_\textrm{tof}$ of the cold Rb cloud.
		Blue diamonds: Release-recapture temperatures of Rb, obtained from the number of recaptured atoms. Systematic errors caused by self-evaporations during the waiting time after recapture lead to overestimated temperatures $T_\textrm{rr}$, increasing with the cloud's temperature $T_\textrm{tof}$.
		Green circles: Release-recapture temperatures of Rb, avoiding systematic errors caused by self-evaporation. 
		The gray line serves as a guide to the eye $T_\textrm{rr} = T_\textrm{tof}$.}
	\label{fig:rr_calibration}
\end{figure}

In our previous work \cite{Spethmann2012} the thermalization of single Cs atoms with a thermal Rb cloud was studied in the collisionless regime \cite{Stamper-Kurn1998}, where the time-scale of thermalization is dominated by the trap frequencies.
Here, the thermalization rate of Cs atoms within the Rb cloud is dominated by the interspecies scattering rate, exceeding the trap frequencies by up to a factor of 100.
This is the reason for the rapid thermalization of the Cs atoms at first interaction with the Rb cloud, giving rise to two distinct fractions of thermalized and non-thermalized atoms.
Since the trap frequencies of the two species almost match, the spatial distributions of the thermalized fraction and the cold cloud are expected to be almost equal as well, consistent with our observations (compare Fig.~\ref{fig:density_variation} and \ref{fig:movie_fraction_analysis}~(b)).
Performing release-recapture thermometry selectively with the thermalized fraction is not possible, since only the total number of remaining atoms from both fractions can be detected after the recapture.
We therefore conduct release-recapture and spatial distribution measurements on the thermometer atoms both with and without the Rb cloud present in the trap as shown in Fig.~\ref{fig:Cs_RR}~(a).
From the spatial distribution measurements in Fig.~\ref{fig:movie_fraction_analysis}, the fractional amount of non-thermalized atoms is determined at the time of release.
With this, an effective release-recapture curve shown in Fig.~\ref{fig:Cs_RR}~(b) of the thermalized Cs fraction is computed as the difference between the two curves, weighted by the sizes of the respective fractions of Cs atoms.

From this effective release-recapture curve, the Rb cloud's temperature is identified as the temperature $T_\textrm{rr}$ for which the corresponding Monte-Carlo simulation for the thermometer atoms fits best.
While the recapture rate for the measurements discussed is around $\unit[50]{\%}$, mainly limited by the initial transfer from the MOT to the dipole potential, it reaches unity in the simulations.
Therefore we rescale the simulated curves using a least-squares fit to the experimental data, with the amplitude as only free parameter.
Around  $T_\textrm{rr}$, the $\chi^2$ value of the fit depends harmonically on the simulation temperature $T_\textrm{sim}$ \cite{Tuchendler2008,Bevington2003} and the uncertainty $\sigma_T$ is given by
\begin{equation}
\chi^2 = \chi^2_0 + \frac{(T_\textrm{sim}-T_\textrm{rr})^2}{\sigma_T^2}
\end{equation}
with $\chi^2_0$ being the $\chi^2$ value of $T_\textrm{rr}$.
The temperatures of the Rb cloud $T_\textrm{rr}$ obtained with single atom thermometry are compared to independently measured time-of-flight temperatures $T_\textrm{tof}$ of the cold cloud itself.
For time-of-flight thermometry, absorption images of the cold cloud are taken for two times of flight $t_i$.
The temperature $T_\textrm{tof}$ is calculated from the 1D widths $\sigma_i$ of a Gaussian fit to the density profiles perpendicular to the horizontal beam, where a harmonic oscillator approximation of the potential holds \cite{Brzozowski2002}
\begin{equation}\label{eq:tof_temperature}
T_\textrm{tof} =\frac{m_\textrm{Rb}}{k_\textrm{B}}\frac{( \sigma_2^2 - \sigma_1^2 )}{(t_2^2 - t_1^2)}
\end{equation}
with $m_\textrm{Rb}$ the Rb mass and $k_\textrm{B}$ the Boltzmann constant.
Small errors $\Delta T_\textrm{tof}\propto \nicefrac{1}{\left(\sigma_2^2 - \sigma_1^2\right)}$ are obtained if one time of flight is large, but for large times of flight absorption imaging yields a low signal to noise ratio. Therefore, we analyze an average of typically 120 images and apply a fringe-removal technique~\cite{Ockeloen2010}.
Fig.~\ref{fig:rr_calibration} shows good agreement between $T_\textrm{rr}$ and $T_\textrm{tof}$ for all measured temperatures, indicating a successful application of single atom thermometry to the cold cloud.

\section{Release-Recapture thermometry on the cold cloud}
\begin{figure}[tb]
	\begin{center}
		\includegraphics{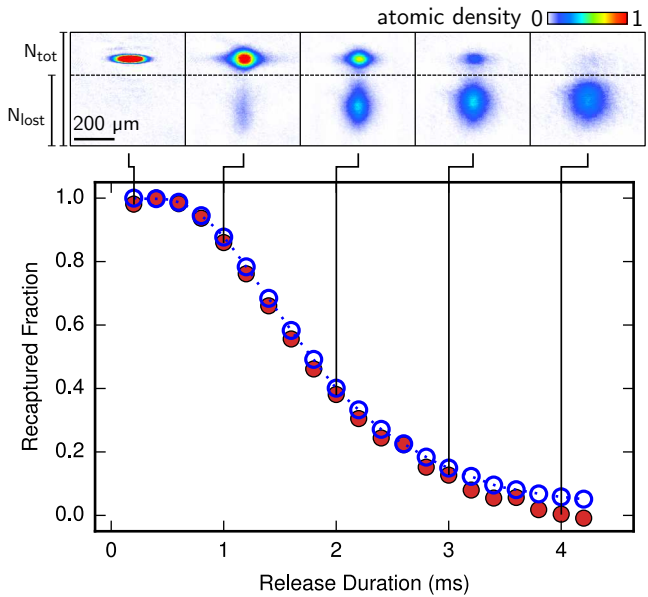}
	\end{center}
	\caption{Release-recapture thermometry on a cold Rb cloud at $T_{\textrm{tof}} = \unit[1.0\pm0.1]{\upmu K}$. Systematic errors due to self-evaporation are avoided by computing the survival rates from the absorption images (top) as $1- \nicefrac{N_{lost}}{N_\textrm{tot}}$, \unit[6]{ms} after the beginning of the release. Experimental survival rates (red dots) best fit a Monte-Carlo simulation (blue circles) at a temperature of $T_{\textrm{rr}} = \unit[1.2]{\upmu K}$. The time-of-flight absorption images show recaptured atoms at the top, while atoms lost during the release appear below. The statistical errors do not exceed the size of the markers.}
	\label{fig:Rb_images_release_duration}
\end{figure}
The presented results of single atom thermometry so far rely on the validity of our Monte-Carlo simulations.
Thus, we verify their reliability by performing release-recapture thermometry with the cold Rb cloud itself and comparing the temperatures obtained to independently measured time-of-flight temperatures.

Starting with laser cooled Rb atoms in the dipole trap, we lower the potential until the cloud is evaporatively cooled to the desired temperature.
Subsequently, release-recapture thermometry is performed analogously to Cs: 
The trap is switched off for a short variable release time and after a waiting time of \unit[50]{ms} the number of remaining Rb atoms is measured with absorption imaging.
The release-recapture temperature $T_\textrm{rr}$ is again obtained by comparison to Monte-Carlo simulations.
A comparison of $T_\textrm{rr}$ to independently measured time-of-flight temperatures $T_\textrm{tof}$ is shown in Fig.~\ref{fig:rr_calibration}:
Strikingly, the release-recapture temperature $T_\textrm{rr}$ matches very well for low temperatures, that correspond to low trap depths and atom numbers, but over-estimates the temperature of the cloud for higher temperatures, which correspond to a deeper potential and higher atom number.
This systematic error is an effect of self-evaporation, caused by the increase in energy during the release.
Short release durations lead to a negligible increase of energy, while long release durations result in a small atomic density after recapture, such that self-evaporation is negligible in both cases.
For intermediate release-times, however, self-evaporation causes additional losses, resulting in a faster decaying release-recapture curve and over-estimation of temperature when compared to the simulations, since self-evaporation is not included in our model.

To avoid this systematic error, the number of atoms lost during release $N_\textrm{lost}$ is measured, which is unaffected by self-evaporation (see Fig.~\ref{fig:Rb_images_release_duration}).
The recaptured fraction of atoms is given as $1-\nicefrac{N_\textrm{lost}}{N_\textrm{tot}}$ with the total number of atoms $N_\textrm{tot}$.
For this method, a comparison to the time-of-flight temperatures $T_\textrm{tof}$ (see Fig.~\ref{fig:rr_calibration}) shows a good agreement, even for higher temperatures that correspond to rapid self-evaporation.
In summary, the release-recapture and time-of-flight thermometry methods are consistent for a wide range of temperatures, which proofs the validity of our Monte-Carlo simulations, and hence of our release-recapture thermometry method.
\\
\section{Conclusion and Outlook}
We have demonstrated the use of single Cs atoms as a thermometer for a cold thermal Rb cloud. 
The cold cloud's temperature is identified by comparing release-recapture measurements with the thermometer atoms to classical Monte-Carlo simulations.
A precise knowledge of the thermometer atoms' spatial distribution proves to be essential for release-recapture thermometry.
Here, we analyze a situation where not all thermometer atoms are thermalized with the cold Rb cloud, and we show how to account for the influence of the non-thermalized fraction.
The validity of the temperatures obtained with single atom thermometry is verified by comparison to independently measured time-of-flight temperatures of the cold cloud.
Also, the reliability of our Monte-Carlo simulations is shown in single-species experiments for both Cs and Rb:
First, we have studied the motion of single Cs atoms, transferred from a MOT to our trap, in absence of the cold Rb cloud.
Our simulations reproduce the observed time dependence of the atoms' spatial distribution well and yield a realistic MOT temperature.
Secondly, we have performed release-recapture thermometry on the cold Rb cloud itself.
The simulations yield temperatures for the cloud consistent with independent results of time-of-flight thermometry.
For this measurement, we have also identified self-evaporation as an important source of systematic errors and presented a way to overcome these perturbations.

Our work thus paves the way to use single Cs atoms as localized and non-perturbing temperature probes.
In particular, the optical lattice has been designed as species selective potential, experienced only by the thermometer atoms \cite{Schmidt2015}.
Thus, a Rb BEC can be quenched into a non-equilibrium state in the trap, while the Cs atoms remain strongly localized in the optical lattice wells.
Measuring the local energy distribution of thermometer atoms immersed in such a Rb system with high time- and position resolution will yield valuable insight into quantum non-equilibrium dynamics as well as system-bath dynamics.

\section{Acknowledgements}
	We thank S. Haupt for initial work on numerical simulations.
	The project was financially supported partially by the European Union via the ERC Starting Grant 278208 and partially by the DFG via SFB/TR49. D.M. is a recipient of a DFG-fellowship through the Excellence Initiative by the Graduate School Materials Science in Mainz (GSC 266), F.S. acknowledges funding by Studienstiftung des deutschen Volkes, and T.L. acknowledges funding from Carl-Zeiss Stiftung.
\newpage
\bibliography{letter.bib}
			
\end{document}